\pdfoutput=1
\documentclass[fleqn,usenatbib]{mnras}

\usepackage{newtxtext,newtxmath}

\usepackage[T1]{fontenc}
\usepackage{ae,aecompl}

\usepackage{graphicx}	
\usepackage{amsmath}	
\usepackage{amssymb}	
 
\title[A dual-beam, dual-camera polarimeter]{Characterization of a dual-beam, dual-camera optical imaging polarimeter}

\author[M. Shrestha et al.]{Manisha Shrestha,$^{1}$\thanks{E-mail: m.shrestha@ljmu.ac.uk}
Iain A. Steele,$^{1}$
Andrzej S. Piascik,$^{1}$
Helen Jermak,$^{1}$
\newauthor Robert J. Smith,$^{1}$
Chris M. Copperwheat,$^{1}$
\\
$^{1}$Astrophysics Research Institute, Liverpool John Moores University, Liverpool Science Park IC2, 146 Brownlow Hill, \\
Liverpool L3 5RF, UK
}

\date{Accepted XXX. Received YYY; in original form ZZZ}

\pubyear{2020}

\begin{document}
\label{firstpage}
\pagerange{\pageref{firstpage}--\pageref{lastpage}}
\maketitle

\begin{abstract}
Polarization plays an important role in various time-domain astrophysics to understand the magnetic fields, geometry, and environments of spatially unresolved variable sources.  In this paper we present the results of laboratory and on-sky testing of a novel dual-beam, dual-camera optical imaging polarimeter (MOPTOP) exploiting high sensitivity, low-noise CMOS technology and designed to monitor variable and transient sources with low systematic errors and high sensitivity.   We present a data reduction algorithm that corrects for sensitivity variations between the cameras on a source-by-source basis.  Using our data reduction algorithm, we show that our dual-beam, dual-camera technique delivers the benefits of low and stable instrumental polarization ($<0.05$\% for lab data and $<0.25$\% for on sky data) and high throughput while avoiding the additional sky brightness and image overlap problems associated with dual-beam, single-camera polarimeters.  
\end{abstract}

\begin{keywords}
instrumentation:polarimeters 
\end{keywords}



\section{Introduction}
The polarization state of light can provide complementary knowledge of the source to that we can infer from other techniques such as photometry, spectroscopy, and imaging. Thus, we are losing valuable information when we ignore the polarization of different astrophysical sources. Polarimetry has proven to be a powerful diagnostic tool for various science cases such as studying environment around stellar sources \citep{Brown_1977,Shrestha_2018}, supernovae remnants \citep{Hoffman_2008}, blazars \citep{Jermak_2017}, gamma ray bursts (GRBs; \cite{Steele_2009,Mundell_2013}), novae \citep{Harvey_2018}, and many more. In recent years polarimetric studies have taken a leading role as a key diagnostic tool of magnetic field strength/order/geometry, geometry of environment surrounding the source, and relativistic plasma dynamics in various time domain sources such as blazars, active galactic nuclei, x-ray binaries, and GRBs. Polarization studies of these time varying sources can provide spatial resolution information not possible by other techniques. 

Optical polarization is traditionally measured via a ratio of fluxes in two or more polarization states obtained consecutively. This approach does not work for rapidly fading targets such as GRBs where an erroneous signal can be introduced due to rapid fluctuation in flux. To address this problem we developed a series of polarimeters (RINGO, RINGO2, and RINGO3) that used a rapidly rotating polaroid analyser \citep{Arnold_2017,Jermak_2017} on the 2.0-m Liverpool Telescope (LT). These instruments have successfully observed various time varying sources. RINGO2 was used to make first ever detection of early-time optical polarization in GRBs \citep{Steele_2009}. \cite{Mundell_2013} observed high polarization of a GRB and its time evolution using RINGO2. RINGO2 and RINGO3 have also been used to produce a comprehensive multi-colour polarization data set of blazar variability \citep{Jermak_2016}.

Multi-colour OPTimised Optical Polarimeter (MOPTOP) is the next generation of polarimeter to replace RINGO3 at the Liverpool Telescope. MOPTOP benefits from the advantages of the RINGO series of instruments such as a fast rotating element to allow high time resolution and the use of fast readout, very low noise cameras. In addition, it uses a unique optical dual-camera configuration to minimize systematic errors and provide the highest possible sensitivity. Dual-beam polarimeters have been built before e.g. \cite{Scarrott_1983}, however they often combine the signal onto a single camera because traditionally the imaging detector has been the most expensive component. This combining causes problems such as source overlap and reduced sensitivity via increased sky background. The recent availability of relatively low cost scientific complementary metal-oxide semiconductor (sCMOS) detectors with high quantum efficiency, very low read noise, and rapid readout, means a dual camera imaging polarimeter becomes feasible. 

In this paper we will characterize a prototype version of MOPTOP and show that a dual-beam, dual-camera polarimeter indeed has better polarimetric accuracy and sensitivity.  We will also describe a data analysis algorithm for dual-camera data that will be available for future MOPTOP science observations. The paper is organized as follows. In Section~\ref{Moptop}, we provide the design of the instrument and its features. Section~\ref{data_reduction} provides the steps of the data reduction pipeline that can be employed to reduce MOPTOP data. We provide details of data observed from lab and on-sky observations in Section~\ref{data}. Then we present the results of these observed data in Section~\ref{results} and provide concluding remarks in Section~\ref{conclusions}.

\begin{figure*}
\includegraphics[width=\textwidth]{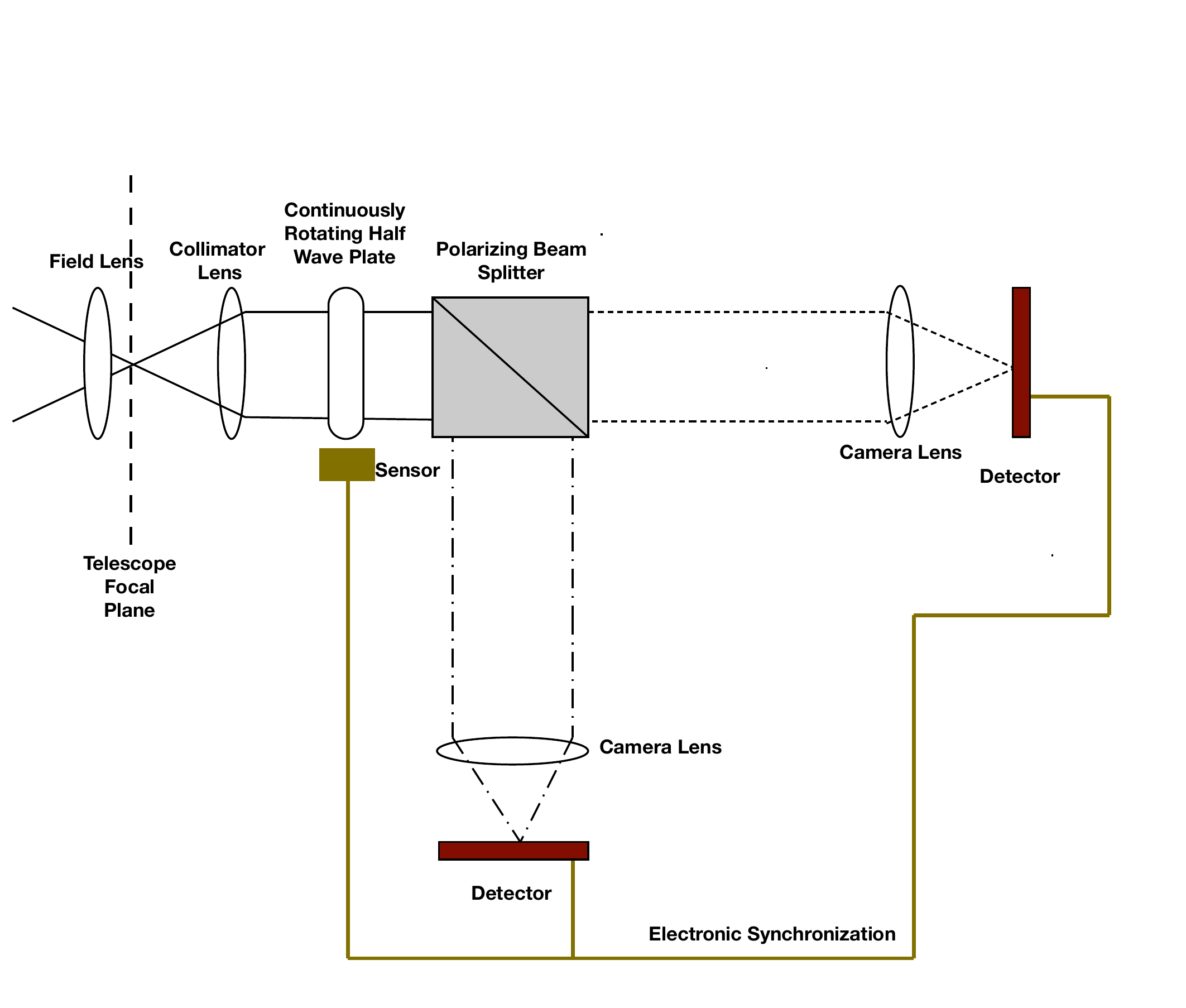}
\caption{Layout of MOPTOP illustrating the design principal and major components from \citet{Jermakspie_2016}. The solid black line represents the incoming light from source. It passes through a field lens, a collimator lens and a continuously rotating half wave plate which modulates the polarization angle of the incoming beam. This beam is split into $s$ and $p$ represented by the dashed and dot-dashed lines. Both polarization states are simultaneously recorded by fast readout sCMOS cameras synchronized to the wave plate angle.}
\label{fig:moptop_sketch}
\end{figure*}

\section{MOPTOP polarimeter}\label{Moptop}

Conceptually, MOPTOP is a fully optimized polarimeter for time-domain astrophysics.  As such it was designed to meet the the following key requirements:

\begin{itemize}
    \item Maximum sensitivity for faint sources.
    \item Low systematic errors allowing measurements of low polarization sources.
    \item Ability to measure a rapidly variable or fading source.
    \item Ability to measure all sources in a wide field of view without source overlap or prior identification.
    \item Ability to measure sources over the widest possible dynamic range of photon counts.
\end{itemize}

The full design concept is explained in detail in \cite{Jermakspie_2016} and the optical design and results of ray-tracing are presented in \cite{jermak-moptop-optics-spie}. MOPTOP is a dual-beam, dual-camera polarimeter which uses two sCMOS cameras per waveband as detectors. The full concept includes an inline filter wheel to provide simultaneous multi-band capability, however, for this paper we present the results of testing a simpler single band version of the instrument.

The advantages of a dual-beam approach to polarimetry have long been known (e.g. \cite{clarke1965}) in that it allows cancellation of the systematic errors that arise in simple single-beam polarimetry. Key to the technique is the use of a half wave plate to modulate the polarization angle of the incoming beam and an analyser system that can simultaneously record the two orthogonal polarization states.  A rotation of the wave plate by 22.5$^\circ$ will rotate the polarization angle of the beam by $45^\circ$, effectively swapping the $q$ and $u$ Stokes parameters.  This means that variations in the polarization response of the instrument can be corrected by a differential technique and gives the consequent reduction in systematic errors.

\begin{figure*}
\includegraphics[width=\columnwidth]{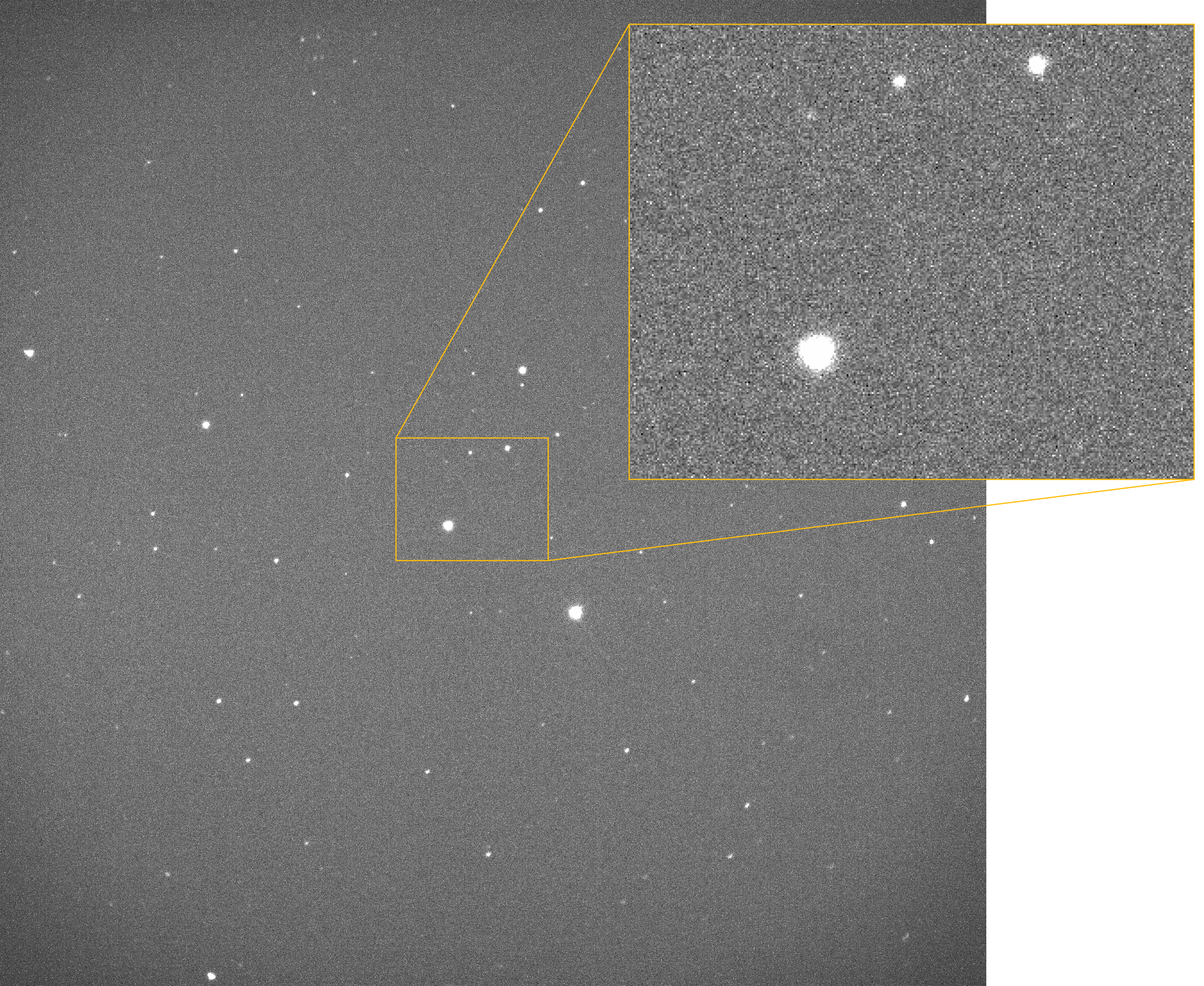}\includegraphics[width=\columnwidth]{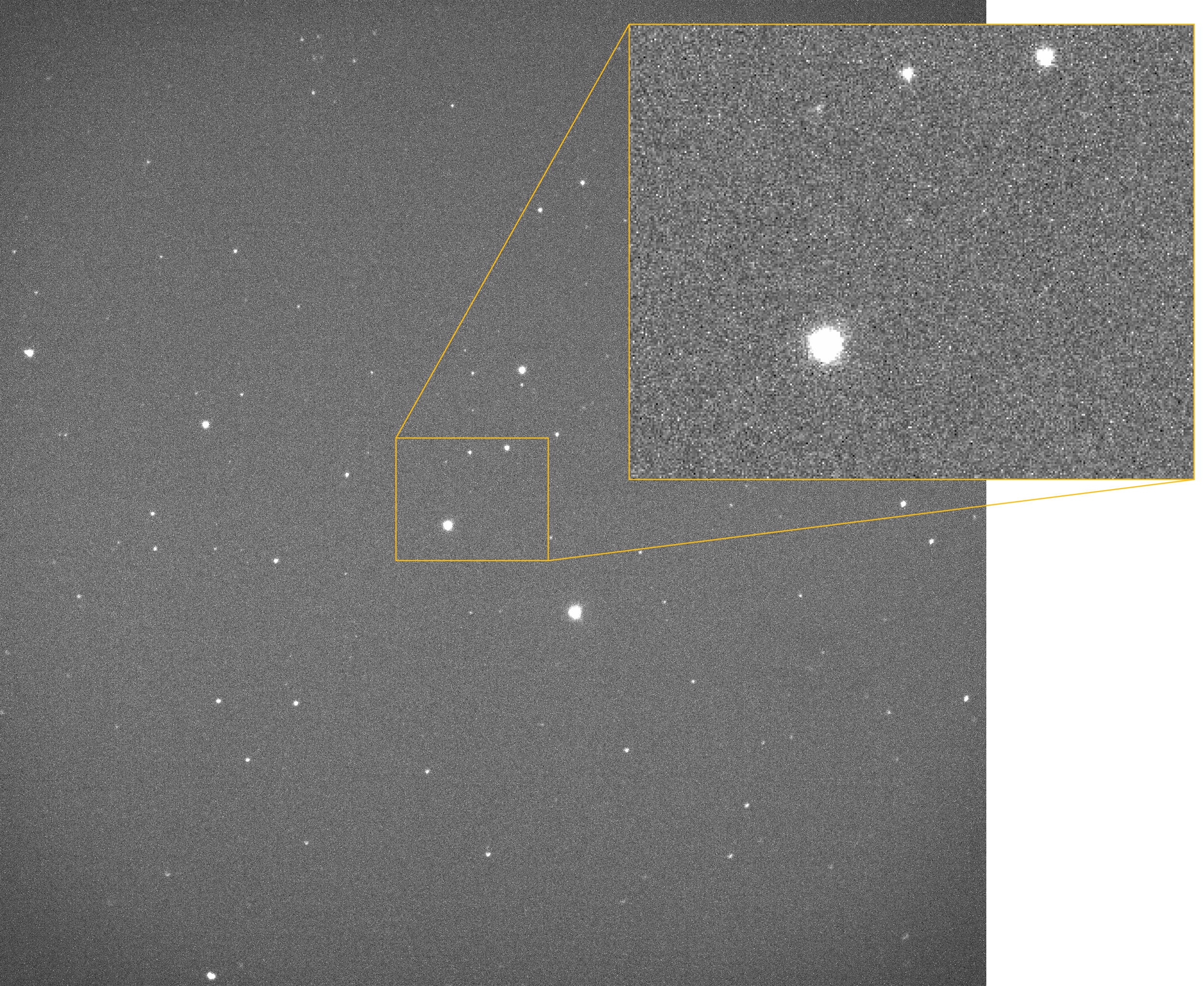}
\caption{Raw image of star field SA111-14 \citep{Landolt_2009} observed by one CMOS camera at two consecutive waveplate positions.  The images show slight vignetting in the field corners due to the optical system.  The zoomed portions of the image show both static dark frame (`hot pixel') and flat field structures and the presence of random telegraph (`popcorn') noise apparent as `hot pixels' that appear in random locations from frame to frame.
}
\label{fig:rawimage}
\end{figure*}

 A sketch of our single band version is presented in Fig.~\ref{fig:moptop_sketch}. The design was implemented using off-the-shelf optical elements, cameras and mechanical components housed within a set of lens tubes. These were then supported by a custom aluminium chassis to interface with the telescope.  Briefly the key elements of the implementation are:
 
\begin{itemize}
    \item An input filter (not shown) that defines the instrument wavelength range.  This is a custom filter comprised of 3-mm Schott GG475 glass cemented to 2-mm Schott KG3 that was originally used in the RINGO 1 and 2 polarimeters.  It gives an approximate wavelength range of 4750 -- 7100 {\AA} (FWHM=2350 {\AA}), and as we showed in \cite{steele-ringo2-2017}, provides a good photometric match to the Sloan $r'$ band. 
    \item An achromatic field ($f=300$-mm) and collimator ($f=150$-mm) lens combination that collimates the output beam from the telescope and places the pupil to coincide with the position of the beam splitter.  Both lenses have an anti-reflection coating with R<1\% per surface between 4000 and 10000 {\AA}.
    \item A Thor Labs achromatic half-wave plate which rotates to modulate the polarization angle of the incoming beam.  This has 98\% transmission (including the effect of anti-reflection coatings) and a retardance within 0.02$\lambda$ of the ideal $\lambda/2$ over the wavelength range defined by the input filter.  The wave plate is mounted in a Physik Instrumente model U-651 rotation stage that in normal operation is continuously rotated at $7.5$ revolutions per minute. 
   
\item A Thor Labs high-throughput polarizing wire-grid beam splitter with 4000-7000 {\AA} anti-reflection coating and a contrast ratio >1000:1 for angles of incidence between 0 and 25$^\circ$ over that wavelength range. This separates the beam into \textit{s} and \textit{p} polarization states to be recorded on two separate cameras.
\item A pair of Nikon SLR camera lenses (50-mm, $f/1.4$) to focus the collimated \textit{s} and \textit{p} beams onto the detectors.
\item A pair of Andor Zyla 4.2 sCMOS detectors with $2048\times2048$ pixels and yielding a field of view of $\sim 7 \times 7$ arcmin. The cameras are specified to have rapid readout (<0.01-s), very low read noise (< 1 $e^-$), and high quantum efficiency ($\sim80$\%).  Readout of the detectors is electronically synchronized to each other and the angle of the wave plate rotation stage using its inbuilt encoder to generate a hardware trigger signal to the cameras \citep{steele-spie-cmos}. An example of two consecutive raw images from one detector is presented in Fig.~\ref{fig:rawimage}.
\item Two compact Intel NUC PCs with Intel i7 processors and M.2 solid state drives running Linux Ubuntu 18.04. The computers are mounted either side of the instrument housing and separately acquire images from each camera over a USB 3.0 interface. One of the PCs also controls the half-wave plate rotator stage and could control a filter wheel in a future upgraded instrument.  
\end{itemize}

\begin{figure*}
\includegraphics[width=\columnwidth]{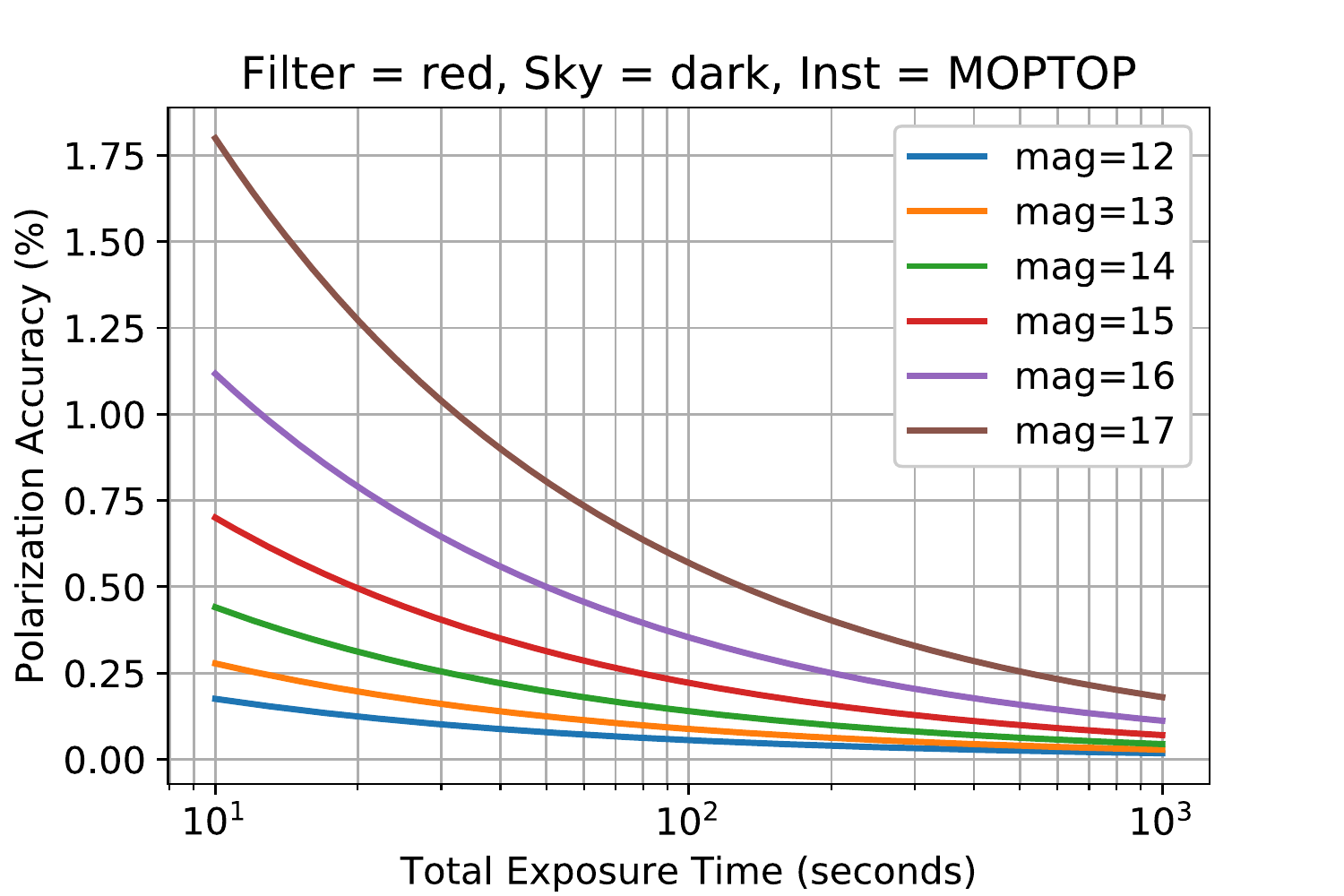}\includegraphics[width=\columnwidth]{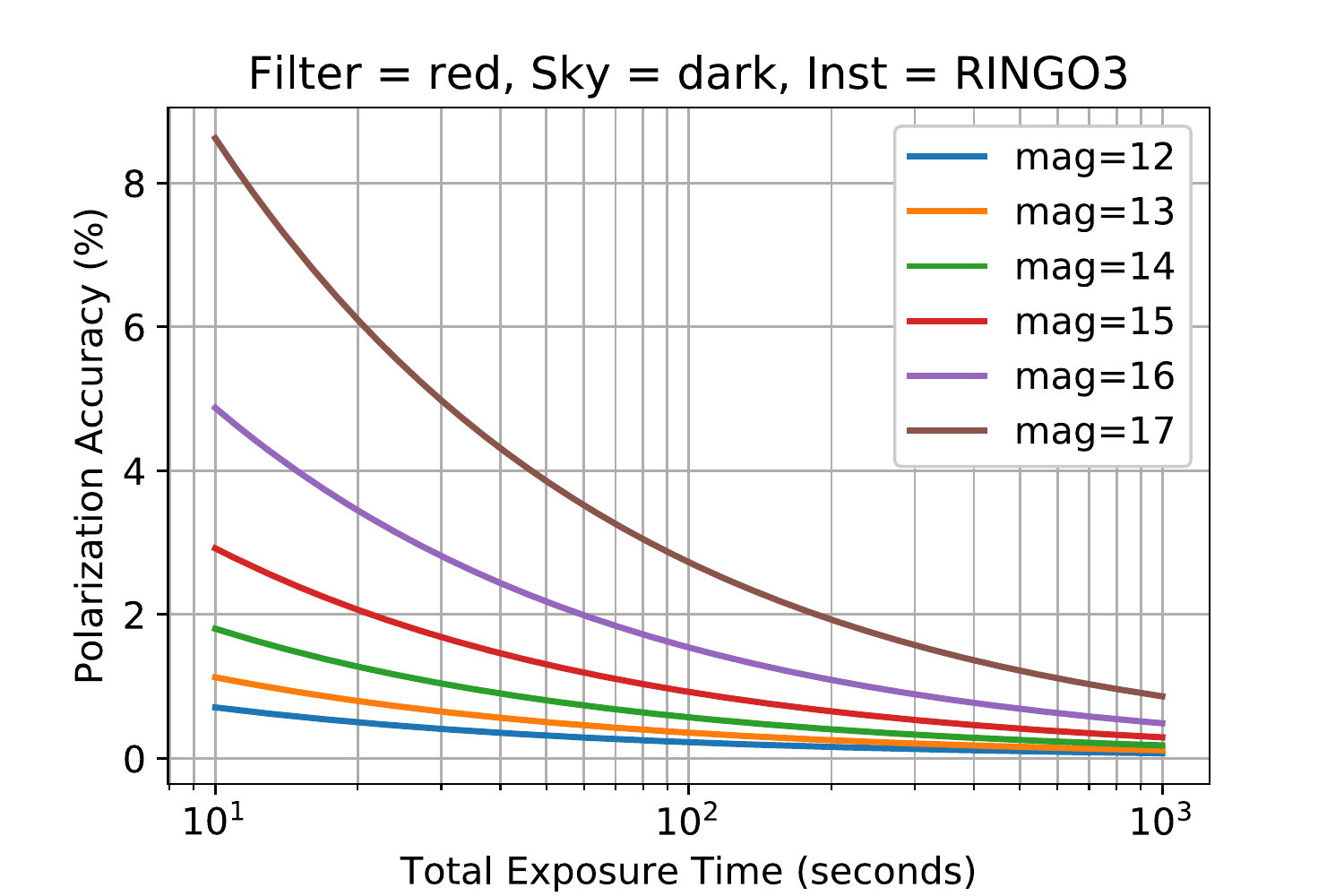}
  \caption{Predicted polarization accuracy with respect to total exposure times for a range of magnitudes represented by different colours. The left figure is for MOPTOP and the right figure is for RINGO3 both for dark sky and magnitude 21.5 in $R$ band filter. The x-axis is in a logarithmic scale for both cases.}
  \label{fig:expo}
\end{figure*}

At low levels of polarization, the sensitivity of a polarimeter will be dominated by systematic effects.  These are hard to predict, however as described above, the dual-beam dual-camera design of MOPTOP aims to minimize these.  As a comparison, the single beam, single-camera RINGO series of polarimeters have suffered systematic effects at the $\sim0.5$\% level \citep{Slowikowska_2016}.  For higher degrees of polarization, polarization sensitivity may be calculated via simple photon statistics, taking into account the sky-brightness and source fluxes (including the system throughputs and detector quantum efficiencies) and detector read-noises.  The results of such a calculation for RINGO3 and MOPTOP are presented in Figure \ref{fig:expo}.  This shows (for example) that a 100 second exposure with MOPTOP on an $R=17$ magnitude object is predicted to attain a polarization accuracy of $\sim0.6$\%, compared to the RINGO3 value of $\sim2.6$\%. There are two principal reasons for this improvement.  Firstly there will be a  doubling of photon throughput from switching from a Polaroid (which  blocks half of the flux) to a beam-splitter arrangement.  Secondly there will be an effective doubling of detector quantum efficiency by moving from EMCCD detectors, which suffer from multiplicative noise in their gain stage \citep{emccd}, to sCMOS.

\section{Data reduction pipeline}\label{data_reduction}


\begin{figure*}
\includegraphics[width = \textwidth]{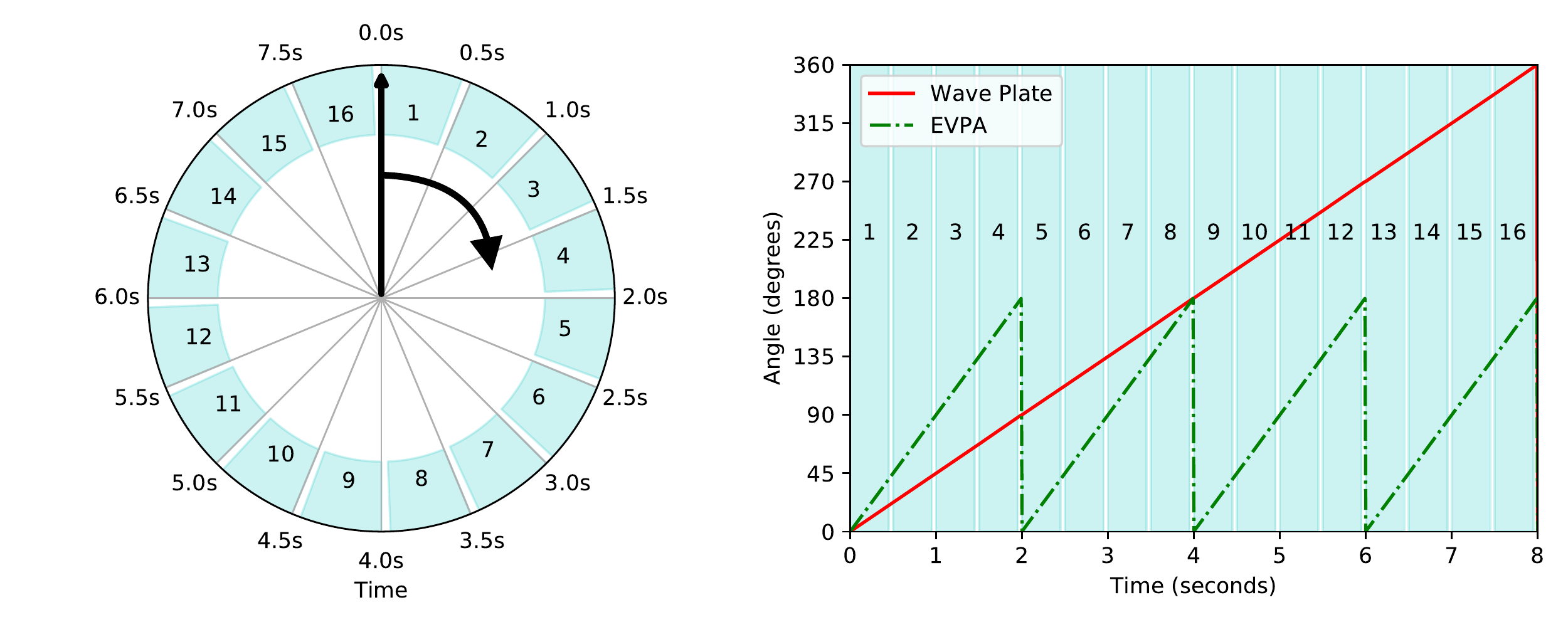}
\caption{ (left) MOPTOP operational principle, showing the sequence of frames taken at different angles of the rotating wave plate as a function of time. Exposure of a frame simultaneously begins on both cameras 0.5-s after the previous one, with the waveplate rotation angle having increased by a total of 22.5 degrees over that time period.  The blue shaded areas indicate the time the shutter is open (duration 0.45-s), with the shorter white regions indicating the 0.05-s readout gaps between frames.  Frames are numbered 1-16, corresponding to the notation introduced in Section \ref{one_camera}.  (right) Waveplate rotation angle and resulting EVPA rotation of the incoming beam as a function of time.  }
 \label{fig:moptop_rot}
\end{figure*}

Data taken with MOPTOP produces 16 different images from each camera, with each image frame corresponding to a $22.5\degr$ rotation of the wave-plate (Fig. \ref{fig:moptop_rot}).  These data can be reduced in two different ways to get polarization values.
Common to both methods is the use of aperture photometry to extract sky-corrected source counts. The extracted counts can then be converted to fractional Stokes parameters \textit{q}-\textit{u} via two different techniques which we call `one-camera' and `two-camera'. In the one-camera technique we use fluxes from 8 frames from one camera to calculate fractional Stokes parameters \textit{q}-\textit{u} following the recipe of \cite{Clarke_2002}. These Stokes parameters can then be used to calculate polarization degree and angle. 

In the two-camera technique we make use of all 16 frames from both cameras. The beam splitter in MOPTOP separates the incoming beam into \textit{s} and \textit{p} polarization states which are recorded on two cameras. Because of the presence of the rotating half wave plate, the difference between the two camera images at the first half wave plate position of $0\degr$ equals the Stokes Q parameter and the difference at half wave plate position of $22.5\degr$ equals the Stokes U parameter. The subsequent third half wave plate position difference yields -Q and the fourth -U. The sum of all the fluxes from four half wave plate position divided by 2 gives the Stokes parameter I. 

\subsection{Flux Extraction}
We perform aperture photometry using the Python package Astropy Photutils \citep{Bradley_2019}. Depending on the source, we use either global background subtraction (when there is a source close to the target) or local background subtraction (when there is no source close to the target). For global background, we estimate the background using a median of the whole image via sigma clipped statistics in Astropy \citep{astropy:2018}. We subtract the global median as the background signal. For local background estimate we use an annulus aperture to estimate the background signal and subtract it from the source. After background subtraction we detect sources above 5 to 15 times the standard deviation of the image signal to noise (SNR) using DAOStarFinder which provides x and y coordinates of the sources in pixel values. 

In order to calculate appropriate aperture size to perform aperture photometry, we check the full width at half-maxium (FWHM) of the source as well as how magnitude and error in magnitude changes with aperture size. We use the combination of these two techniques i.e. an aperture size between 2 to 3 times FWHM and that produces good SNR as the appropriate aperture size. This aperture size around the sources is used to perform aperture photometry and calculate the counts from all the sources. At the end of the routine we have counts values for the sources with a certain threshold SNR for all the 16 frames required for polarization calculations. Along with counts we also calculate the error in these counts which is the root mean square sum of background noise and Poisson noise of the source \citep{Bradley_2019}. The values we get from photometry are used to calculate the polarization of the sources.

\subsection{One-camera technique} \label{one_camera}
In our one-camera technique we implement the procedure outlined by \cite{Clarke_2002}.  This is also the procedure we used for RINGO, RINGO2, and RINGO3 data reduction (e.g. \cite{Jermak_2016}).
In this method the data from one camera for 8 consecutive rotor frames are used to calculate the Stokes parameters. The observed photoelectron-counts for an object in each frame are given by $m_n$ where $n$ represents the frame number (Figure \ref{fig:moptop_rot}).   These photoelectron-counts are then used to calculate the quantities $S_1$, $S_2$, and $S_3$: 
\begin{align}
   & S_1 = m_1+m_2+m_3+m_4+m_5+m_6+m_7+m_8, \\
  & S_2 = m_1+m_5+m_2+m_6,\\
  & S_3 = m_2+m_6+m_3+m_7.
   \label{eq:counts}
\end{align}
 $S_1$, $S_2$, and $S_3$ can then be used to calculate the Stokes parameters \textit{q} and \textit{u} using the \cite{Clarke_2002} equations:
 \begin{equation}
    q = \frac{Q}{I} = \pi \left ( \frac{1}{2}-\frac{S_3}{S_1} \right),
    \label{eq:stokes_q}
\end{equation}{}
\begin{equation}
    u = \frac{U}{I} = \pi \left ( \frac{S_2}{S_1}-\frac{1}{2} \right).
    \label{eq:stokes_u}
\end{equation}{}
Using the Stokes parameters \textit{q} and \textit{u}, we can then calculate the polarization degree (\textit{p}) and polarization angle ($\psi$):
\begin{equation}
    p = \sqrt{q^2+u^2},
    \label{eq:pol}
\end{equation}
\begin{equation}
    \psi = \frac{1}{2} \arctan\left ( \frac{u}{q} \right).
    \label{eq:pa}
\end{equation}
In Eq.~\ref{eq:pa} we take care to determine the correct quadrant for $\psi$.

Errors can be calculated using the photoelectron-count errors as follows:
\begin{align}
   & S_{1e} = \sqrt{m_{1e}^2+m_{5e}^2+m_{2e}^2+m_{6e}^2+m_{3e}^2+m_{7e}^2+m_{4e}^2+m_{8e}^2 },\\
&    S_{2e} = \sqrt{m_{1e}^2+m_{5e}^2+m_{2e}^2+m_{6e}^2},\\
 &   S_{3e} = \sqrt{m_{2e}^2+m_{6e}^2+m_{3e}^2+m_{7e}^2}.
\end{align}
where subscript `e' denotes the error in corresponding measurements.  We then do error propagation in Equations~\ref{eq:stokes_q} and \ref{eq:stokes_u} to calculate the error in corresponding Stokes parameters:
\begin{equation}
     q_{e} = \pi \sqrt{\left (\frac{s_{3e}}{s_1}\right)^2 + \left( \frac{s_{1e} s_3}{s_1^2} \right)^2},
     \label{eq:single_q_e}
\end{equation}
\begin{equation}
      u_{e} = \pi \sqrt{\left (\frac{s_{2e}}{s_1}\right)^2 + \left( \frac{s_{1e} s_2}{s_1^2} \right)^2}.
      \label{eq:single_u_e}
\end{equation}
    
Finally we use the Monte Carlo technique of \cite{steele-ringo2-2017} which is based on \cite{Simmons_1985} to correct for polarization bias and calculate the error in polarization degree and position angle from the errors in $q$ and $u$.

\subsection{Two-Camera technique} \label{two_camera}
The one-camera technique described in Section~\ref{one_camera} is similar to the technique we use with RINGO3. The novel technique of the MOPTOP dual-beam polarimetry is properly utilized when we reduce the polarization values using a two-camera technique. As described in Section \ref{data_reduction}, we can take the difference between two cameras at various half-wave plate angles to get Stokes Q, U, -Q, and -U. To do this we need to ensure the observed counts from the two different cameras have the same effective sensitivity. This is unlikely to be true by default, as the two beams and cameras will have slightly different flat-field and quantum efficiency characteristics.  

For each individual source in a set of images, we define the quantity $F$ (`the sensitivity factor') as the ratio of counts between camera 2 and camera 1 for an unpolarized source at a certain location on the camera.  We then define $m_1$ and $n_1$ as the observed counts from camera 1 and camera 2 respectively for half-wave plate angle $0\degr$ (Fig. \ref{fig:moptop_rot}). We then define $c_1$ and $d_1$ as the counts after correction for the difference in sensitivity.  Hence $c_1 = m_1$ and $d_1 = \frac{n_1}{F}$. We can calculate Stokes Q as $c_1 - d_1$ and Stokes I as $c_1+d_1$, hence we can calculate fractional q as:

\begin{equation}
    q = \frac{Q}{I} = \frac{c_1-d_1}{c_1+d_1}.
    \label{eq:fraq_c}
\end{equation}

Substituting to the counts observed
\begin{equation}
    q = \frac{Q}{I} = \frac{m_1-n_1/F}{m_1+n_1/F}.
    \label{eq:fraq_m}
\end{equation}

Similarly the counts of camera 1 and camera 2 at $22.5\degr$ can be written as $c_2$ and $d_2$ respectively which can be converted to observed counts as $c_2 = m_2$ and $d_2 = \frac{n_2}{F}$
And the counts of camera 1 and camera 2 at $45\degr$ can be written as $c_3$ and $d_3$ respectively which can be converted to observed counts as $c_3 = m_3$ and $d_3 = \frac{n_3}{F}$.

The difference in this counts give us Stokes -q
\begin{equation}
    -q = \frac{Q}{I} = \frac{m_3-n_3/F}{m_3+n_3/F}.\\
    q = \frac{n_3-m_3/F}{n_3+m_3/F}.
    \label{eq:franq_m}
\end{equation}

We can equate Equation~\ref{eq:fraq_m} and the negative of Equation~\ref{eq:franq_m}, which gives us 

\begin{equation}
    F =\sqrt{\frac{n_1 n_3}{m_1 m_3}}.
\end{equation}

We can therefore calculate the sensitivity factor F for all the 8 pairs of subsequent frames. We use each F factor calculated for a pair to correct for the counts corresponding to those frames. Finally these corrected counts are utilized to calculate the Stokes \textit{q} (Equation~\ref{eq:fraq_m}) and \textit{u} via:

\begin{equation}
    u = \frac{U}{I} = \frac{m_2-n_2/F}{m_2+n_2/F}.
    \label{eq:frau_m}
\end{equation}

We then take the average of two consecutive Stokes \textit{q} and \textit{u} to get our final Stokes \textit{q} and Stokes \textit{u} from a set of four images. These \textit{q} and \textit{u} values calculated via the two camera technique can be used in Equation~\ref{eq:pol} and \ref{eq:pa} to calculate the polarization degree and position angle respectively.

We use the error propagation in Equations~\ref{eq:fraq_m} and \ref{eq:frau_m} to calculate the error in Stokes $q$ and $u$ respectively: 

\begin{equation}
    \begin{split}
    & q_{e} = \sqrt{n_{1e}^2+(((m_3/(2n_1n_3m_1))^2 \times(1/(n_1n_3m_1)^2)\times}\\  &\sqrt{((n_{1e}/n_1)^2+(n_{3e}/n_3)^2+(m_{1e}/m_1)^2))+(m_{3e}/m_3)^2)}.
    \end{split}
    \label{eq:two_q_err}
\end{equation}
 
 \begin{equation}
     \begin{split}
    & u_{e} = \sqrt{n_{2e}^2+(((m_4/(2 n_2 n_4 m_2))^2 \times(1/(n_2 n_4 m_2)^2)\times}\\  &\sqrt{((n_{2e}/n_2)^2+(n_{4e}/n_4)^2+(m_{2e}/m_2)^2))+(m_{4e}/m_4)^2)}.
    \end{split}
    \label{eq:two_u_err}
 \end{equation}
 
Where subscript `e' is the error in the corresponding measurement.

Again we use the Monte Carlo technique as per Section ~\ref{one_camera} to calculate the error in polarization degree and position angle.

\section{Data}\label{data}
To study MOPTOP's initial performance, data were collected in the lab (see Section \ref{labtests}) and polarized and unpolarized standard stars were observed on sky (see Section \ref{obs}) using the LT.

\subsection{Laboratory testing}\label{labtests}
The instrument was setup in the lab to test the basic polarization performance without the influence of the telescope optics.  We aimed to simulate three different light sources: unpolarized ($\sim 0\%$),partially polarized ( $4-5\%$), and  highly polarized ($\sim 100\%$). 

In order to simulate an unpolarized light source we experimented with both incandescent bulb and LED light sources, eventually arriving at array of five white-light LEDs arranged with their intrinsic polarizations in a radial pattern.  This was used to illuminate a domestic sucrose sugar-cube which acted as a polarizaton scrambler before the light was fed via a 10-m length of multi-mode optical fibre which further scrambled the polarization and produced a small, bright source at its output that acted as an artificial star.  To create a partially polarized source, the output from the artificial star was then reflected at $45^\circ$ from a front-surface aluminium mirror.  This should produce a polarization of $4-5$\% \citep{Cox_1976}.  Finally an off-the shelf Polarioid sheet was used to produce a highly polarized source option.

These sources were observed using MOPTOP and the polarization of the sources was calculated following the data reduction pipeline described in Section~\ref{data_reduction}. Results are presented in Section~\ref{results}. 

\subsection{Observations on sky}\label{obs}
MOPTOP was mounted on the 2.0-m fully robotic Liverpool Telescope \citep{Steele_2004} in the Roque de Los Muchachos Observatory, La Palma, Spain for three days of observations from 2019 Sept 08 to 2019 Sept 10. MOPTOP on LT yielded a field of view of $7 \times 7 $ arcmin ($4\times$ greater than the RINGO series of instruments).  The sCMOS detector gains were set at $0.55 e^-/ADU$ with a median read noise of $0.9 e^-$.

For the three nights of observations, we selected various standard polarized and unpolarized stars whose polarization degree is available in literature. For the better understanding of instrument performance, we varied various parameters like the Cassegrain rotator angle, star field, and varying magnitude of the targets. The list of observed targets are presented in Table~\ref{tab:general}.

\begin{table*}
	\centering
	\caption{Targets observed by MOPTOP on 2019 Sept 8 -- 10 at LT.}
	\label{tab:general}
	\begin{tabular}{lcccc} 
		\hline
		Name & RA (J2000) & Dec. (J2000) & R mag & Comments \\
		\hline
		BD +32 3739& 20 12 02.11 & +32 47 43.5 & 9.3 & Unpolarized Standard\\
		BD +28 4211 & 21 51 11.07 & +28 51 51.8 & 10.5 & Unpolarized Standard\\
		HD 212311 & 22 21 58.55 & +56 31 52.8  & 8.12 & Unpolarized Standard\\
		HILT 960 & 20 23 28.53  & +39 20 59.1 & 10.6 & Polarized Standard\\
		VICyg 12& 20 32 40.94 & +41 14 26.2 & 11.5 & Polarized Standard\\
		HD 204827 & 21 28 57.70 & +58 44 24.0 & 7.94 & Polarized Standard\\

		KIC 8462852 & 20 06 15.4 & +44 27 24.7 & 11.75 & \cite{Boyajian_2016}\\

		\hline
	\end{tabular}
\end{table*}

\section{results}\label{results}
Following an initial confirmation of camera linearity via observations of the standard field SA111-SF2 \citep{Landolt_2009}, we analyzed our laboratory and on sky data using the data reduction pipeline described in Section~\ref{data_reduction}. The following subsections present our polarization results and show that our dual-beam, dual-camera polarimeter produces better polarization accuracy and sensitivity than our previous single-camera, single beam designs. 
 



\subsection{One-camera technique vs two-camera technique}

First as a proof of concept we checked how the two-camera technique compared to the one-camera technique. We reduced all the 16 frames of data for each camera and implemented the data reduction pipeline as described in Section~\ref{data_reduction}. From each technique, we get four different Stokes \textit{q} and \textit{u} parameters. We calculated polarization degree and error in polarization degree. 

\begin{figure*}
\includegraphics[width=\columnwidth]{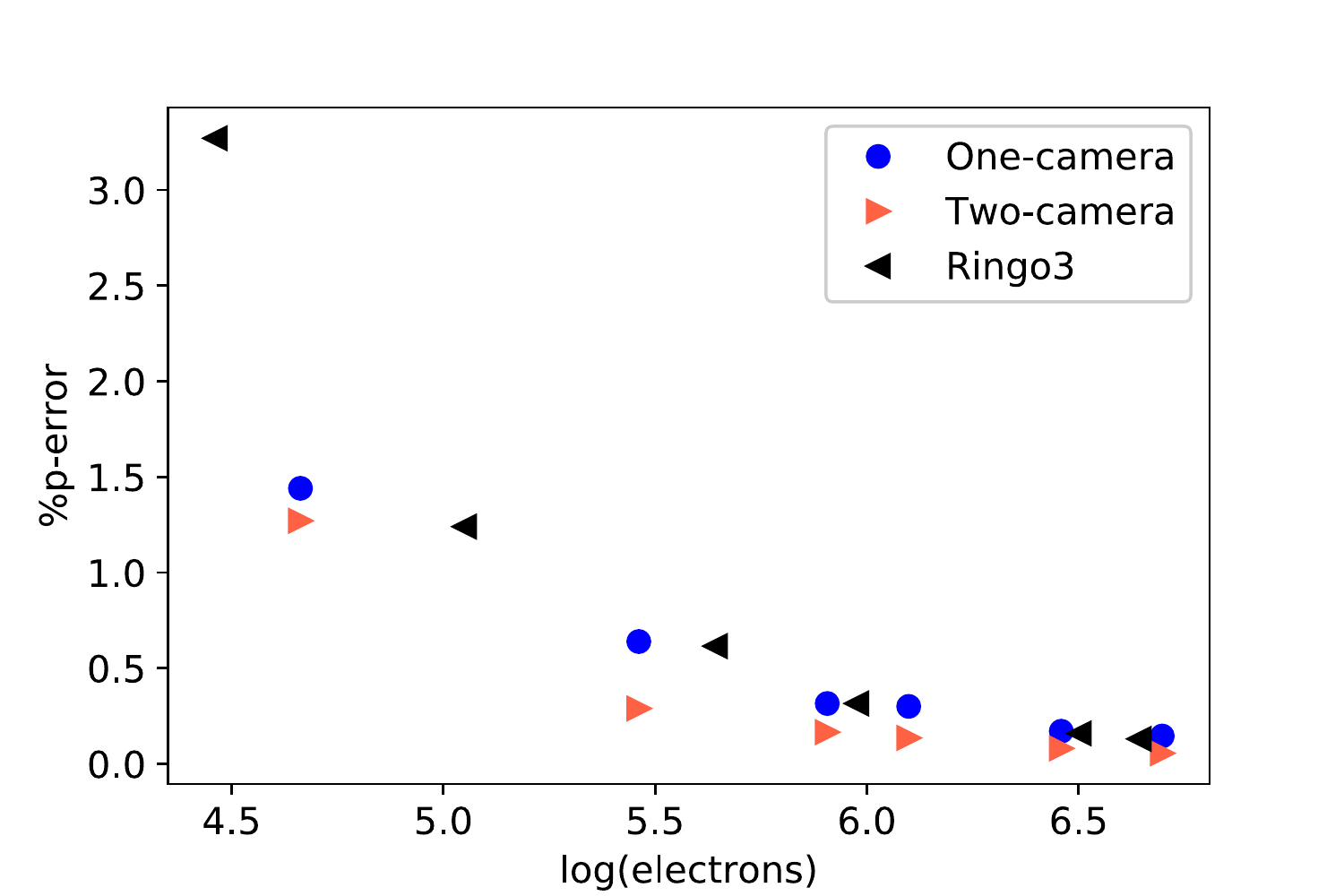}\includegraphics[width=\columnwidth]{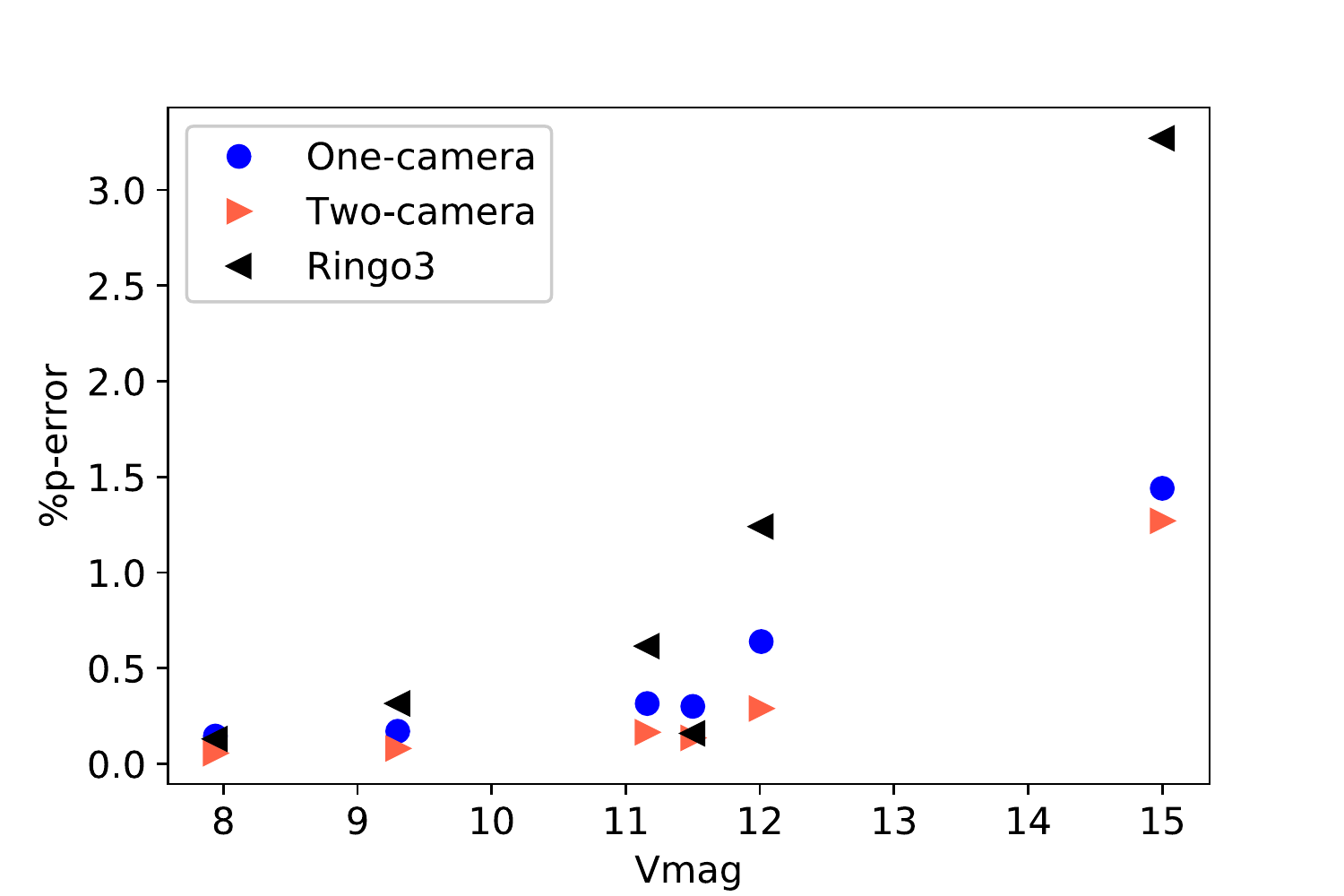}
  \caption{Calculated formal errors (derived from Eq. \ref{eq:single_q_e}, \ref{eq:single_u_e}, \ref{eq:two_q_err}, \ref{eq:two_u_err}) in polarization degree for equal length integration times with MOPTOP and RINGO3 of the Landolt field SA111-SF2.  Blue circles represent results from MOPTOP one-camera technique, red triangles are for the two-camera MOPTOP technique, and black triangles for data from RINGO3.    The left hand figure is plotted with respect to the total number of photo-electrons recorded and shows similar polarization accuracy for RINGO3 and the single camera analysis of MOPTOP.  The advantage of the two-camera MOPTOP technique is clear.  The right hand figure shows polarimteric error as a function of source magnitude, and demonstrates the throughput advantage of MOPTOP even in one camera mode.
}
  \label{fig:perr}
\end{figure*}

Figure~\ref{fig:perr} shows how error in polarization degree varies with electron counts (left) and magnitude (right) for the one-camera and two-camera techniques for MOPTOP observations and RINGO3 observations. We can see that the two-camera technique has the lowest error in polarization degree compared to the other two techniques which is expected as the two-camera technique makes use of all the photons observed whereas the one-camera and RINGO3 techniques only use half of the photons detected by the instrument. We can also compare the results in Fig.~\ref{fig:perr} from observations with results in Fig.~\ref{fig:expo} from simulations. We find that the observed error in polarization degree is close to the simulated error in polarization degree. For example, the $12^{th}$ magnitude star with $\sim 6$ seconds exposure time has $0.289 \%$ error in polarization degree from observations and from simulations the expected error in polarization degree is $0.226\%$. 

\begin{figure}
\includegraphics[width=\columnwidth]{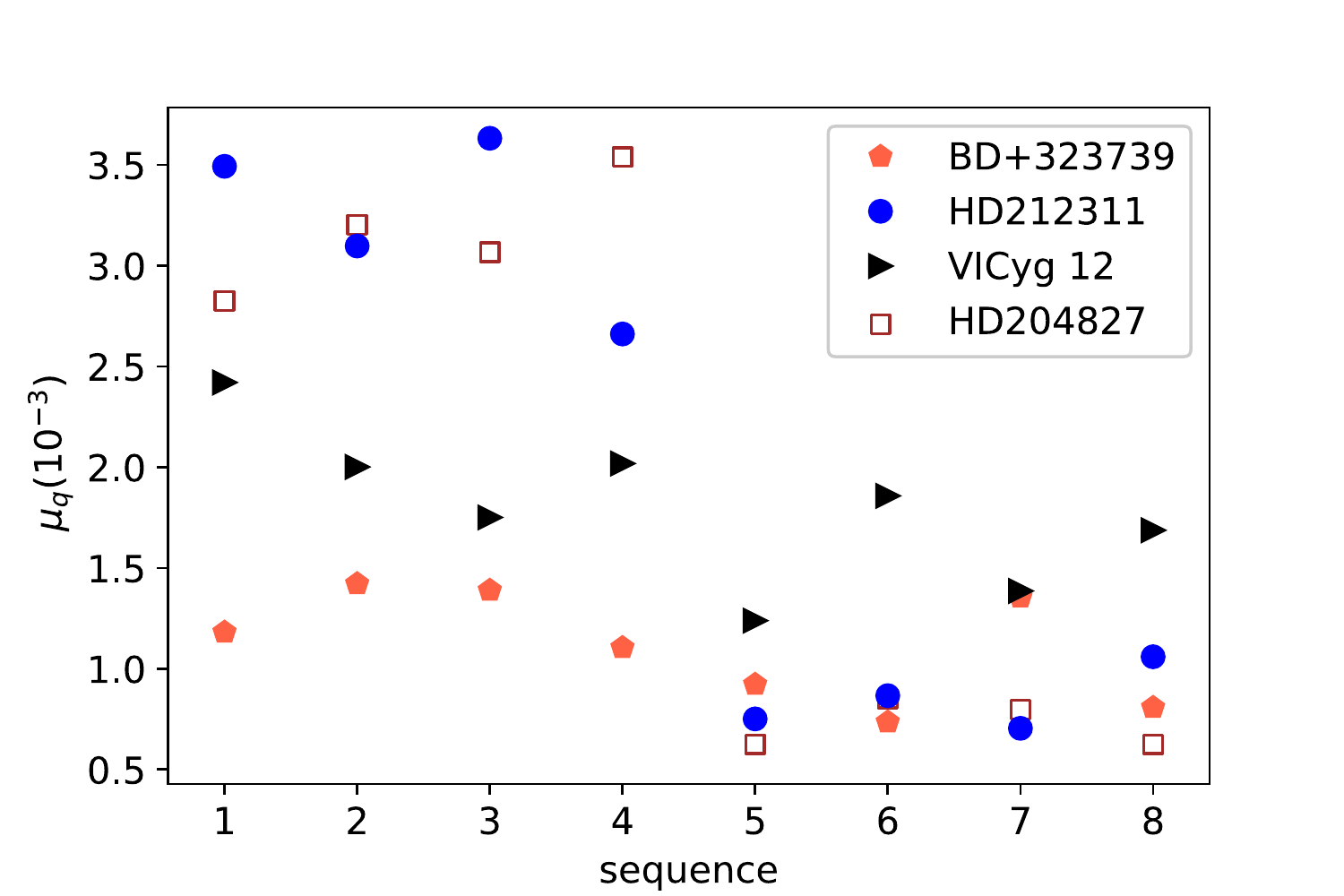}
  \caption{Standard deviation in Stokes \textit{q} with respect to different sequences of observations for different stars observed. First four sequences are for one-camera technique and last four are for two-camera technique. }
  \label{fig:std}
\end{figure}

Another way to compare the two techniques was to look at standard deviation in different Stokes \textit{q} values for the one-camera and two-camera techniques for different sources. Figure~\ref{fig:std} presents standard deviation in Stokes \textit{q} for eight different values (first four are from one-camera technique and last four are from two-camera technique) we get from MOPTOP observations. We can see for all the sources observed, the two-camera technique has less scatter in Stokes \textit{q} compared to the one-camera technique. Thus, showing that the two-camera technique has less systematic error compared to the one-camera technique. As expected we see the same behaviour for Stokes \textit{u}.


These comparisons show that the two-camera technique produces better results with less error in polarization degree and less scatter in Stokes \textit{q} and \textit{u} compared to the one-camera technique. For the rest of the results sections we therefore only present the two-camera technique for clarity unless stated otherwise. 

\subsection{Laboratory data results}
\begin{figure*}
\includegraphics[scale=0.93]{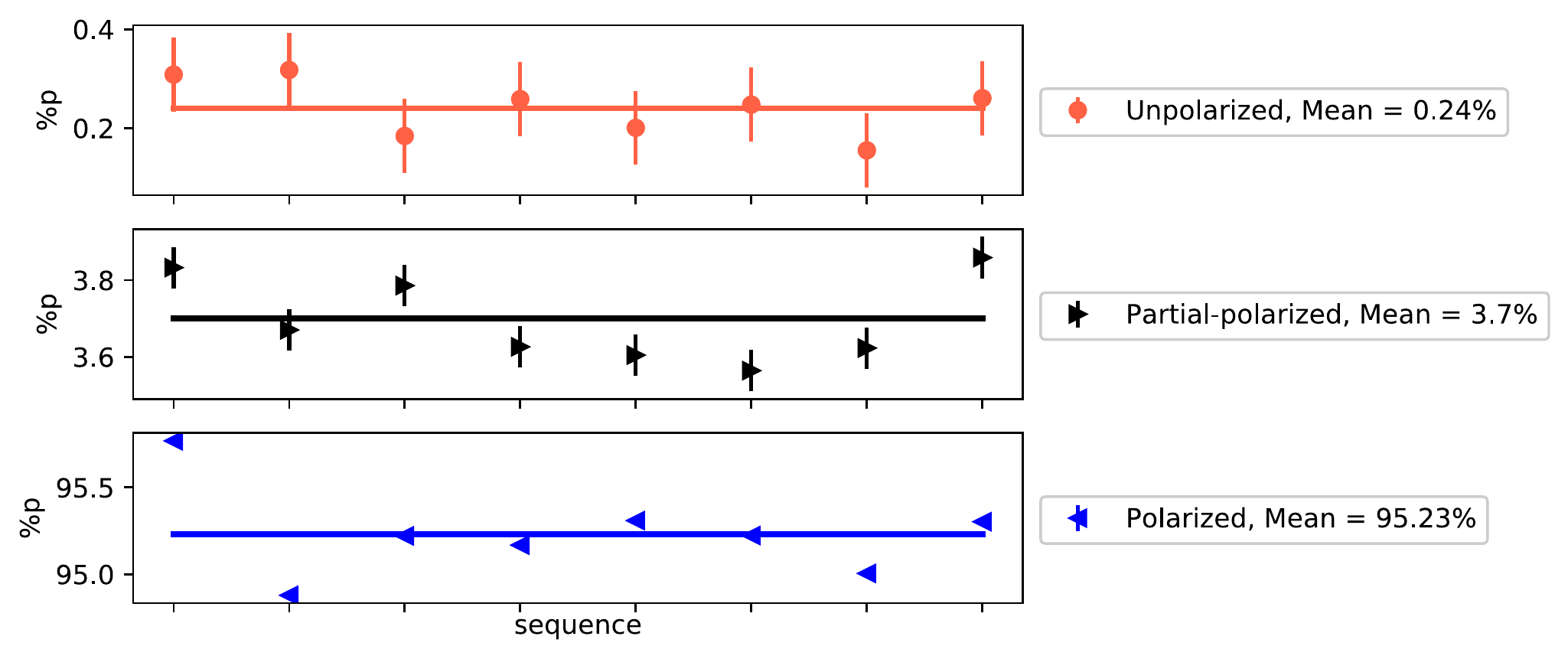}
  \caption{Polarization degree for different set of laboratory data. The three panels are for simulated light in the lab of unpolarized light ($\sim 0 \%$), partially polarized light ($4-5 \%$), and fully polarized light ($\sim 100 \%$) for top panel, middle panel, and bottom panel respectively. The points represent data points from observation and the solid line represents the mean of the data points. Error in polarization degree is shown by the vertical solid line. The label consists of mean value for each panel. }
  \label{fig:labdata}
\end{figure*}

Our laboratory test data was analyzed using the two-camera technique to calculate polarization degree to characterize the instrument before on-sky observations. Fig.~\ref{fig:labdata} includes the results of the observed polarization degree for various sources. For our unpolarized source, the mean of all the observed polarization degrees is $0.24\%$ and the standard deviation is $0.054$. While it is possible that the laboratory source is not fully unpolarized, this shows that the intrinsic polarization of the instrument must be less than that value.

For the partially polarized source we found a mean of $3.7\%$ with a standard deviation of $0.106$.  This is in line with expectations ($4-5\%$), and the low standard deviation indicates that systematic errors are likely to be low.

Finally for our fully polarized source, we get a mean of $95.23\%$ and standard deviation of $0.245$.  Given the source is likely not-fully polarized due to the limited contrast ratio of off-the-shelf Polaroid, we can use this result to establish an upper limit to any instrumental depolarization of $<5$\% of the measured value. 

Overall these results show that instrument is working as expected and introduces a very low level of instrumental polarization and depolarization in the case where there is no background noise and no additional polarization from mirrors of the telescope.

\subsection{Unpolarized standard stars}\label{unpolarized}
The first three stars in Table~\ref{tab:general} are unpolarized standards we observed during the observation run. The archival data show that these stars are unpolarized and the values are presented in Table~\ref{tab:obsvalues}. BD+323739 and BD+284211 were observed with exposure time per frame 0.46 seconds and HD212311 was observed with exposure time per frame 0.2 seconds. We varied Cassegrain rotor angles ranging from $-90\degr$ to $+90\degr$ with an increment of $22.5\degr$. We observed these stars for two nights in a row. In Table~\ref{tab:obsvalues}, we present the observed polarization degree compared to catalogue values for these stars. 

\begin{figure*}
\includegraphics[scale=0.9]{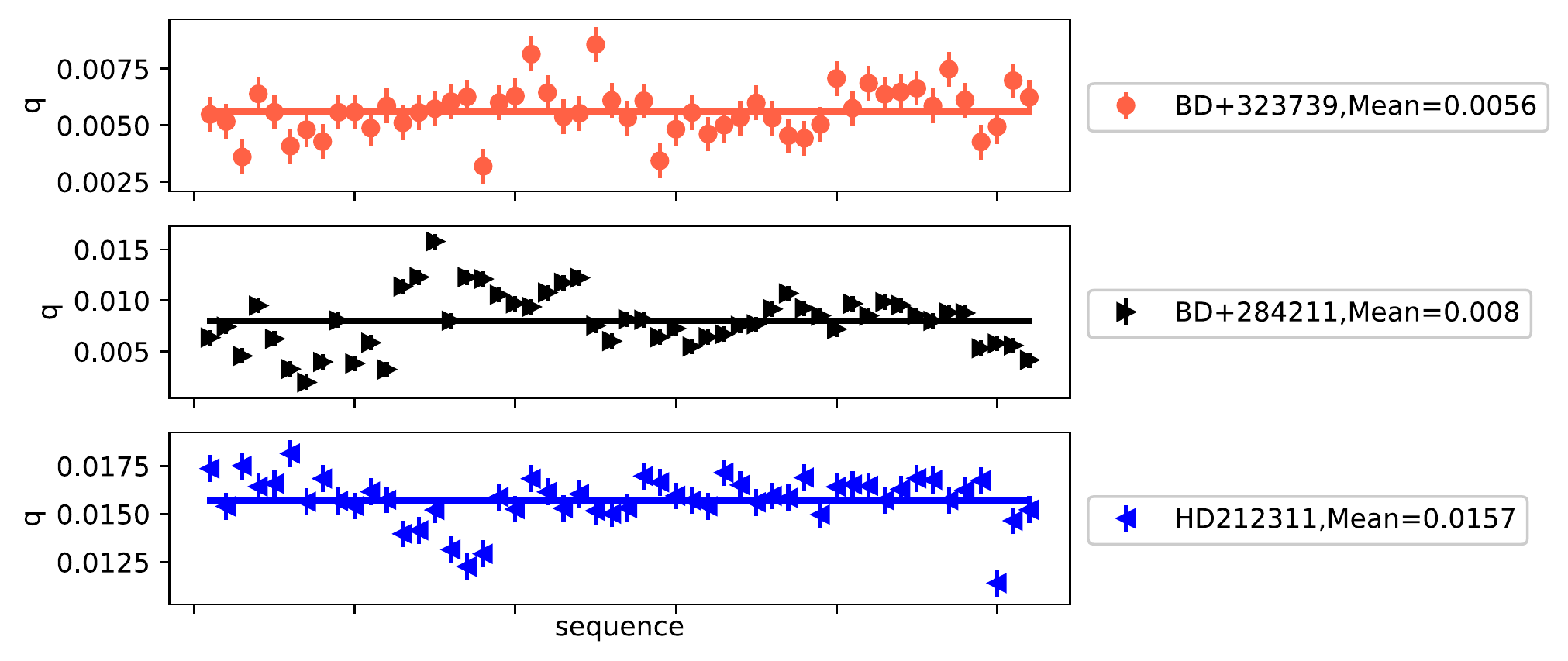}
  \caption{Stokes q vs sequence of observed data for unpolarized standard stars BD +32 3739, BD +28 4211, and HD 212311 top, middle, and bottom panels respectively before we correct for instrumental polarization. The points represent data points from observation and the solid line represents the mean of the data points. The vertical line represents the error bar.
  }
  \label{fig:qzero}
\end{figure*}

\begin{figure*}
\includegraphics[scale=0.9]{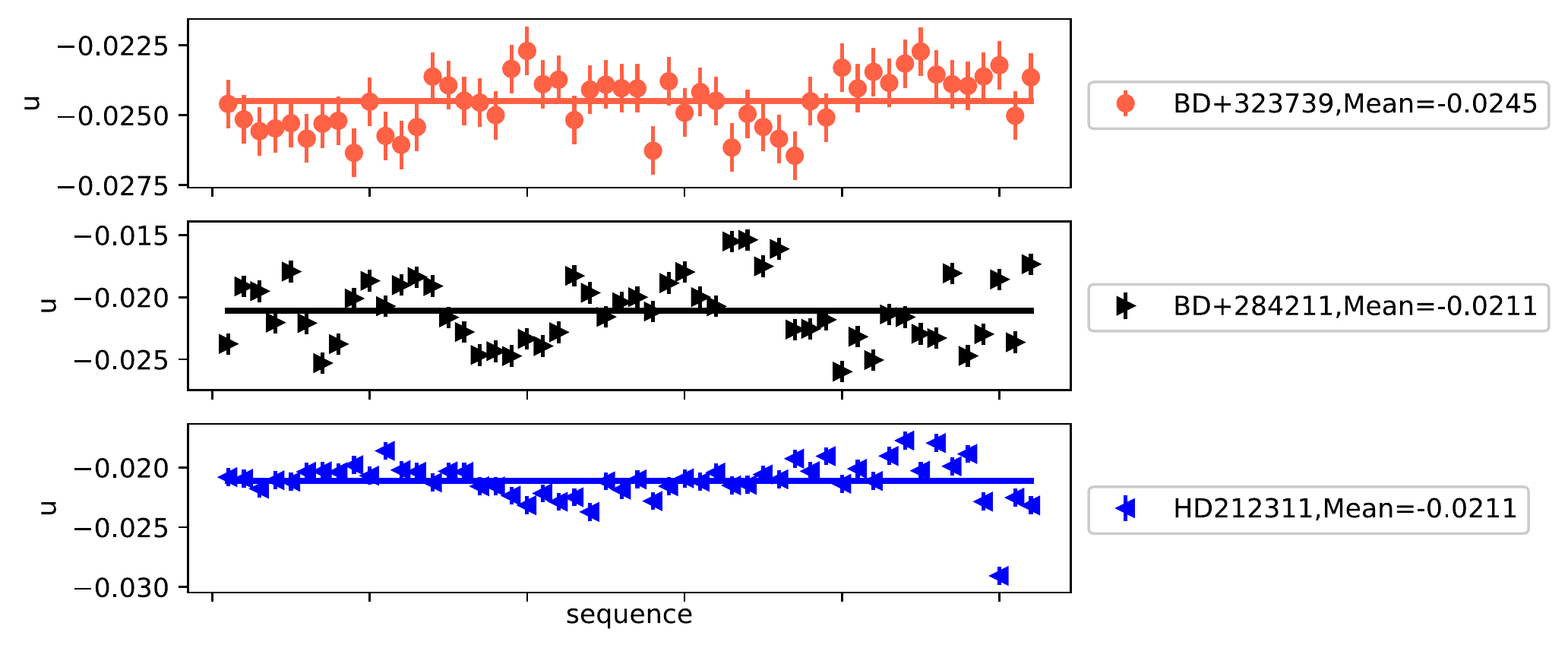}
  \caption{Stokes u vs sequence of observed data for unpolarized standard stars BD +32 3739, BD +28 4211, and HD 212311 from top, middle, and bottom panels respectively before we correct for instrumental polarization. The points represent data points from observation and the solid line represents the mean of the data points. The vertical line represents the error bar.
  }
  \label{fig:uzero}
\end{figure*}

For the observed data, we used the two-camera technique to reduce the data and obtained Stokes \textit{q}, \textit{u} and errors in \textit{q} and \textit{u}. We find there is a smaller scatter in Stokes \textit{q} and \textit{u} for these stars from MOPTOP observations compared to those seen from RINGO3 observations \citep{Slowikowska_2016}. We present the results of observed unpolarized standard stars in Figs.~\ref{fig:qzero} and~\ref{fig:uzero}. For BD+323739, BD+284211, and HD212311 , we find $q_{avg} = 0.0056,0.0080, 0.01570$ respectively and standard deviation $q_{std} = 0.00105, 0.00273, 0.00126$ respectively. For Stokes \textit{u} the averages are, $u_{avg} = -0.0245,-0.0211,-0.0211$ respectively and standard deviations are $u_{std}=0.00097, 0.00265, 0.001706$ respectively as shown in Figs.~\ref{fig:qzero} and~\ref{fig:uzero}. These values are calculated for all the observed data taken during observational run and averaged over four different Stokes parameters we get from the two-camera technique. As these are unpolarized standard stars, we expect Stokes \textit{q} and \textit{u} to be zero. We see that the mean of Stokes $q$ depends on exposure time since this will be related to the mean wave plate angle during an exposure. Thus, we get different average Stokes \textit{q} for HD212311 which had half the exposure time of other two unpolarized standard stars. Hence the average from each star is considered as the additional polarization introduced by the instrument, dependent on exposure time.

\begin{table*}
	\centering
	\caption{Polarimetric archival data of the polarized and unpolarized standard stars along with data from MOPTOP observation for R band before correcting (Uncor) for instrumental polarization and after corrected (Cor). Errors are calculated using the Monte Carlo technique described in Section~\ref{data_reduction}. Polarimetric archival data from \citet{Schmidt_1992} for the unpolarized (in V band) and polarized (in R band) standard stars. The catalogue polarization value for KIC 8462852 is from \citet{Steele_2018} for R band.}
	\label{tab:obsvalues}
	\begin{tabular}{lcccc} 
		\hline
		Name & Cat \%p & Obs \%p Uncor & Obs \%p Cor \\
		\hline
		BD +32 3739 & 0.025 $\pm$ 0.017 &2.513 $\pm$ 0.080& -\\
		BD +28 4211 & 0.054 $\pm$ 0.027 &2.245 $\pm$ 0.080&-\\
		HD 212311 & 0.034 $\pm$ 0.021 &2.620 $\pm$ 0.075& -\\
		
		HILT 960 & 5.210 $\pm$ 0.029  &-&- \\
		VICyg 12& 7.893 $\pm$ 0.037 &6.840$\pm$0.129& 6.965 $\pm$0.129\\
		HD 204827 & 4.893 $\pm$ 0.029 & 4.673 $\pm$0.064 &4.755$\pm$0.064 \\
		
		KIC 8462852 & 0.6 $\pm$ 0.1  &2.273$\pm$0.280& 0.750 $\pm$ 0.280  \\

		\hline
	\end{tabular}
\end{table*}

\subsection{Polarized standard stars}\label{polarized}
The next three stars in Table~\ref{tab:general} i.e. Hilt 960, VICyg~12, and HD~204827 are polarized standard stars we observed using MOPTOP on the LT during its commissioning run. These targets are relatively highly polarized and have good archival data as presented in Table~\ref{tab:obsvalues}. Hiltner~960 and VICyg~12 were observed with exposure time 0.46 seconds per frame and HD~204827 was observed with exposure time 0.2 seconds per frame.

\begin{figure*}

\includegraphics[width=\textwidth]{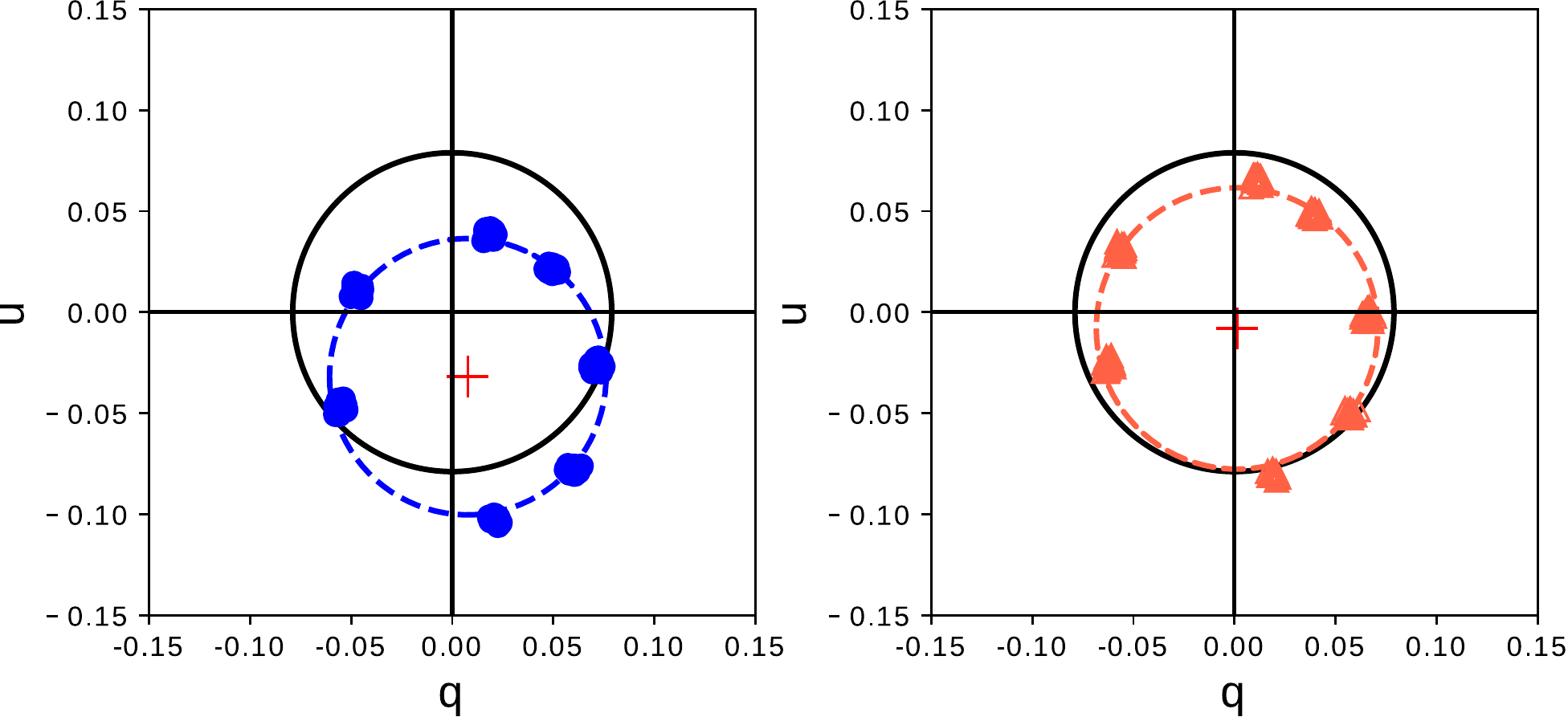}
  \caption{Stokes q vs Stokes u for polarized standard star VI-Cyg12 before correction for instrumental polarization on left and after correction for instrumental polarization on right. The red triangle and blue circle points represent the observed values for different Cassegrain rotor positions, dashed line is a best fit of the observed data, red cross represents the center of the best fit, solid black lines represent the q=0 and u=0 lines for reference. And the solid black circle is the catalogue value of VICyg~12. Error bars are smaller than the points. }
  \label{fig:qu-VICyg}
\end{figure*}

\begin{figure*}

\includegraphics[width=\textwidth]{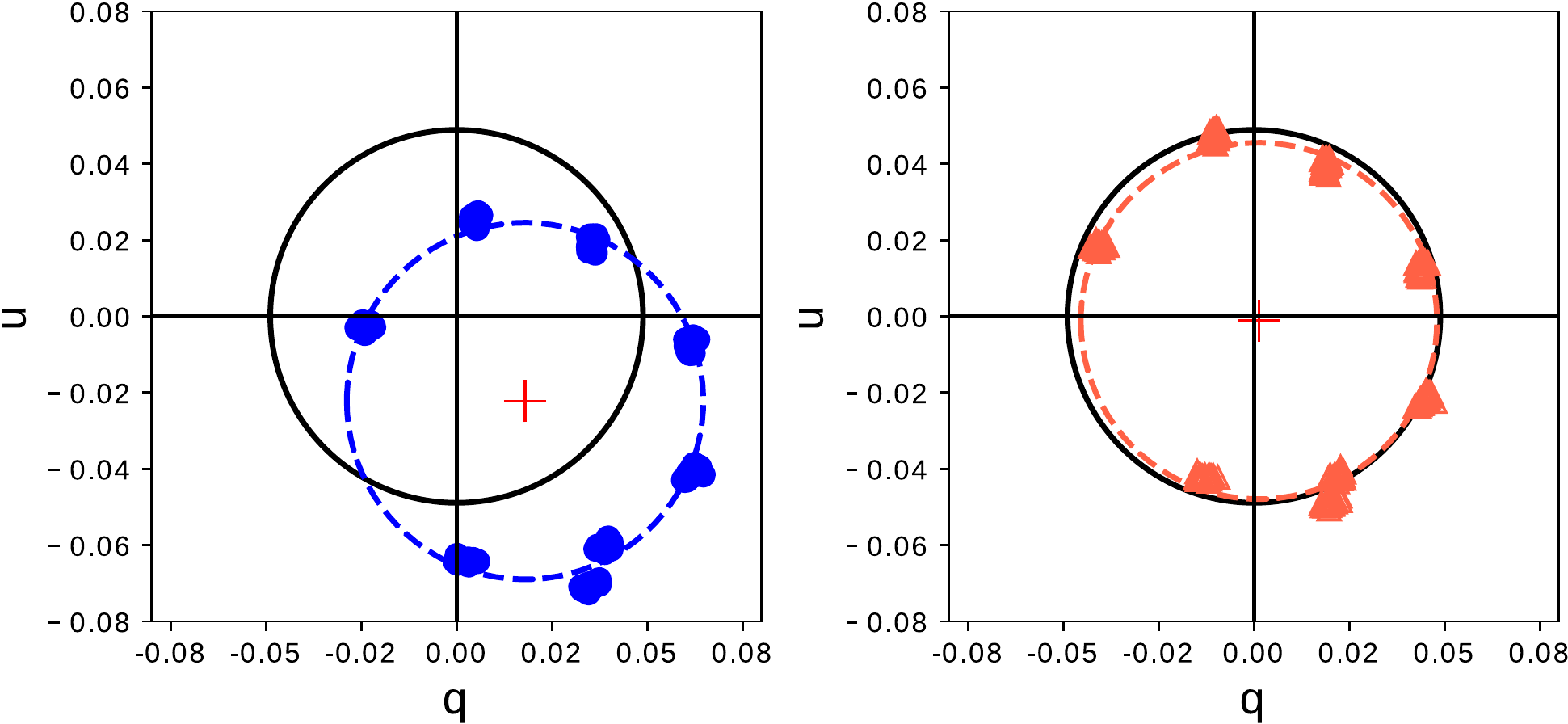}
  \caption{Stokes q vs Stokes u for polarized standard star HD~204827 before correction for instrumental polarization on left and after correction for instrumental polarization on right. The red triangle and blue circle points represent the observed values for different Cassegrain rotor positions, dashed line is a best fit of the observed data, red cross represents the center of the best fit, solid black lines represent the q=0 and u=0 lines for reference. And the solid black circle is the catalogue value of HD~204827. Error bars are smaller than the points.  }
  \label{fig:qu-hd2048}
\end{figure*}

We used the data reduction pipeline as discussed in Section~\ref{data_reduction} to reduce the polarized standard star data. Results for only two of the polarized standard stars observed are presented in Figs.~\ref{fig:qu-VICyg} and \ref{fig:qu-hd2048} because we observed Hiltner~960 at different Cassegrain rotor position for the first night and the data quality for that day is not on the par with other observations. Blue points represent the observed Stokes \textit{q}-\textit{u} values for different Cassegrain rotor positions before correcting for instrumental polarization. Dashed blue lines represent the best fit for observed data and the solid black line is the catalogue value. The red cross shows the centre of the best fit circle for the observed data points. For corrected values, the circle should centre on the origin. In both cases, the red triangle represent data after they were corrected for instrumental polarization using unpolarized standard stars. For red points in Fig.~\ref{fig:qu-VICyg}, we use results from BD~+28 4211 ( $q_{inst} = 0.008$ and $u_{inst} = -0.021$) as our instrumental polarization because both have same exposure time. For Fig.~\ref{fig:qu-hd2048} we use results from HD~212311 ( $q_{inst} = 0.0157$ and $u_{inst} = -0.0211$) as instrumental polarization for correction because both have same exposure time. 

After the correction, the centre of the data moves closer to the origin and the best fit circle of dashed red line is more circular and more closely matched to the solid black circle created using catalogue values. The observed value for HD~212311 after correction for instrumental polarization is close to the catalogue values as seen in Fig.~\ref{fig:qu-hd2048}. However for VICyg~12 the match is not as good between observed and catalogue values as can be seen in Fig.~\ref{fig:qu-VICyg}. For VICyg~12 we see apparent significant ($\sim15$\%) depolarization in observed values compared to the catalogue value.  The difference between these two objects means we cannot comment definitively on on-sky instrumental depolarization based on this data set alone.  However we note that we did not see evidence for strong instrumental polarization in our laboratory data.  We also note an apparent lower than catalogue polarization for VICyg~12 in other literature such as \cite{Simmons_1985} and \cite{Arnold_2017}. Thus, we believe it is more likely this behaviour is due to polarization variability in the source rather than an instrumental effect. In Table~\ref{tab:obsvalues} we present the observed polarization degree compared to archival data.

Another way to estimate instrumental polarization is to check the distance between the origin and the red cross. From this analysis method we get $q_{inst} = 0.029$ and $u_{inst} = -0.019$ for VICyg~12 observations compared to $q_{inst} = 0.008$ and $u_{inst} = -0.021$ from unpolarized standard star BD~+28 4211. We can see discrepancy in $q_{inst}$ and $u_{inst}$ between the two methods here, which we can see on the right panel of Fig.~\ref{fig:qu-VICyg}. For shorter exposure time, we get $q_{inst} = 0.0183$ and $u_{inst} = -0.023$  for HD~204827 compared to $q_{inst} = 0.0157$ and $u_{inst} = -0.0211$ from unpolarized standard HD~212311. The match is better in this case and can be seen in left panel of Fig.~\ref{fig:qu-hd2048}. 

\subsection{Effect of Cassegrain Rotator Angle}

Finally, to characterize the instrument in more detail, we observed the three unpolarized and three polarized standard stars at different Cassegrain rotor angles.  We present results for three unpolarized standard stars BD+323739, BD+284211, and HD212311 in Figs.~\ref{fig:qzero} and \ref{fig:uzero} for observed Stokes \textit{q} and \textit{u} using two-camera technique. Mean values of these observations are presented in the label. We observed these stars at different Cassegrain rotor positions ranging from $-90 \degr$ to $90\degr$ with $22.5\degr$ intervals. In Figs.~\ref{fig:qu-VICyg} and \ref{fig:qu-hd2048} we present $q$-$u$ plots for observed data before and after correcting for instrumental polarization, left and right respectively for polarized standard stars VI-Cyg12 and HD 204827. We found that using the average of unpolarized standard star observations (with the same exposure time and Cassegrain rotor angle as the polarized standard observations) gave the best results. Hence, for future observations we recommend observing unpolarized standard stars with the same exposure time and Cassegrain rotor angles as science targets to estimate instrumental polarization.

\section{Conclusions}\label{conclusions}
In this paper, we have studied the performance of a prototype dual-beam, dual-camera polarimeter (MOPTOP) in laboratory testing and mounted on the Liverpool Telescope.  We have described and tested a data reduction procedure that corrects for intrinsic sensitivity differences between each camera on a source-by-source basis.   Analysis of observations shows that the technique of using two-cameras in a dual-beam polarimeter produces better polarization accuracy for the same exposure time as compared to a single-beam, single-camera polarimeter (such as RINGO3) as shown in Fig.~\ref{fig:expo} and lower error in polarization degree as shown in Fig.~\ref{fig:perr}.  There is better stability in the data from the two-camera technique compared to the one-camera technique as shown in Fig.~\ref{fig:std} for all the targets observed on-sky. Thus, we recommend observers use the two-camera technique for the data reduction of future MOPTOP observations.

In lab data we find instrumental polarization $<0.25\%$ with a stability of $\sim 0.05\%$ and a maximum degree of instrumental depolarization of $<5\%$ of the measured value. On-sky testing showed an instrumental polarization of $2.4\%$ caused by the telescope fold mirror with a stability of $\sim 0.25\%$.

In summary, we have shown observationally that a dual-beam polarimeter implemented using dual-cameras performs better than single beam methods. It also avoids the image overlap and sky noise issues of dual-beam, single camera polarimeters.

\section{Acknowledgements}
We thank our anonymous referee for thoughtful comments that have greatly improved the paper. LT is operated on the island of La Palma by Liverpool John Moores University at the Spanish Obervatorio del Roque de los Muchachos of the Instituto de Astrofisica de Canarias, with financial support from the UK Science and Technologies Facilities Council (STFC). Financial support for the development of MOPTOP, MS and ASP was also provided by STFC. We acknowledge with thanks the variable star observations from the AAVSO International Database contributed by observers worldwide and used in this research. This research made use of Photutils, an Astropy package for detection and photometry of astronomical sources (\cite{Bradley_2019}).




\bibliographystyle{mnras}
\bibliography{moptop} 







\bsp	
\label{lastpage}
\end{document}